\documentclass[preprint,10pt,5p]{elsarticle}

\usepackage{amssymb}
\usepackage{mathrsfs}

\usepackage{graphicx}
\usepackage{dcolumn}
\usepackage{bm}
\usepackage{mathrsfs}
\usepackage{multirow}   
 
\usepackage{lineno}

\usepackage{atlasphysics} 

\usepackage{lmetdefs} 

\begin{document}

\journal{Physics Letters B; CERN-PH-EP-2011-023}

\begin{frontmatter}

\title{Search for high-mass states with one lepton plus missing transverse momentum \\ 
in proton-proton collisions at $\sqrt{s}= 7 \tev$ with the ATLAS detector }

\author{The ATLAS Collaboration}

\date{March 24, 2011}

\begin{abstract}
The ATLAS detector is used to search for high-mass states, such as heavy charged gauge bosons (\wp, \wstar),
decaying to a charged lepton (electron or muon) and a neutrino. 
Results are presented based on the analysis of \pp\ collisions at a center-of-mass energy of
7~\tev\ corresponding to an integrated luminosity of 36~\ipb.
No excess beyond standard model expectations is observed.
A \wp\ with sequential standard model couplings is excluded at 95\% confidence level for masses below 1.49~\tev, 
and a \wstar\ (charged chiral boson) for masses below 1.35~\tev.
\end{abstract}


\end{frontmatter}


Although the standard model (SM) of strong and electroweak
interactions is remarkably consistent with particle physics observations to date,
the high-energy collisions at the CERN Large Hadron Collider provide new opportunities
to search for physics beyond it.
One extension common to many models is the existence of additional heavy gauge
bosons~\cite{pdg:2010}, the charged ones commonly denoted \wp.
Such particles are most easily searched for in their decay to a charged lepton (either electron or muon)
and a neutrino.

In this letter, 7~\tev\ \pp\ collision data collected with the ATLAS detector during 2010 and corresponding to a total
integrated luminosity of 36~\ipb\
are used to supplement current limits~\cite{d0:Wprime, cdf:Wprime, cdf:Wprime2010, cms:wprime2010, cms:wpmu2011mar}
on \xbr\ (cross section times branching fraction) as a function of \wp\ mass.
A lower limit on the mass of a \wp\ boson in the Sequential Standard Model (SSM)~\cite{ssm} is also reported.
In this model, the \wp\ has the same couplings to fermions as the SM \w~boson and
thus a width which increases linearly with \wp\ mass.

Limits are also established for \wstar, the charged partner of the chiral bosons described
in~\cite{wzstar}. Theoretical motivation for such bosons is provided in~\cite{wzstar_motivate}.
The anomalous (magnetic-moment type) coupling of the \wstar\ leads to kinematic distributions
significantly different from those of the \wp.
To fix the coupling strength, a model with total and partial decay widths equal to those of the SSM \wp\ 
with the same mass is adopted~\cite{wzstar_refmod2}.

The analysis presented here identifies candidates in the electron and muon channels,
sets separate limits for \mbox{\wpse} and \wpsmu, and derives combined limits assuming flavor independence.
The kinematic variable used to identify the \wps\ is the transverse mass
\begin{linenomath}
\begin{equation}
\mt = \sqrt{ 2 \pt \met (1 - \cos \varphi_{l\nu})}
\end{equation}
\end{linenomath}
which displays a Jacobian peak that, for \wpl, falls sharply above the resonance mass.
Here \pt\ is the lepton transverse momentum, \met\ is the magnitude of the missing transverse momentum (\mettext),
and $\varphi_{l\nu}$ is the angle between the \pt\ and \mettext\ vectors.
Throughout this letter, transverse refers to the plane perpendicular to the colliding beams, longitudinal
means parallel to the beams, \mytheta\ and \myphi\ are the polar and azimuthal angles with respect to the
longitudinal direction, and pseudorapidity is defined as $\eta = -\ln(\tan(\theta/2))$.

The main background to the \wp\ and \wstar\ signals comes from the high-\mt\ tail of SM \wlnu\ decay.
Other backgrounds are \z\ bosons decaying into two leptons where one lepton is not reconstructed,
\w\ or \z\ decaying to \tauleps\ where the $\tau$ subsequently decays to an electron or muon, and
diboson production. These are collectively referred to as the electroweak (EW) background.
In addition, there is a background contribution from \ttbar\ production which is most important for the lowest \wps\ masses
considered here where it constitutes about 20\% of the background after final selection.
Other QCD background sources, where a light or heavy hadron decays semileptonically or a jet is misidentified as an electron,
are estimated to be at most 3\% of the total background
(with the uncertainty on this estimate less than 10\% of the total background level).

The ATLAS detector~\cite{atlas:detector} has three major components: the inner (tracking) detector, 
the calorimeter and the muon spectrometer.
Charged particle tracks and vertices are reconstructed with silicon pixel and silicon strip detectors
covering $|\eta|<2.5$ and transition radiation detectors covering $|\eta| < 2.0$,
all immersed in a homogeneous 2~T magnetic field provided by a superconducting solenoid.
These are surrounded by a finely-segmented, hermetic calorimeter system that covers \mbox{$|\eta| < 4.9$}
and provides three-dimensional reconstruction of particle showers.
It uses liquid argon for the inner electromagnetic
compartment followed by a hadronic compartment based on scintillating tiles in the central region ($|\eta| < 1.7$)
and additional liquid argon for higher $|\eta|$.
Outside the calorimeter, there is a muon spectrometer with air-core toroids providing a magnetic
field, whose integral averages about 3~Tm.
Three stations of drift tubes and cathode strip chambers provide precision measurements and resistive-plate and thin-gap
chambers provide muon triggering capability and measurement of the $\varphi$ coordinate.

Most of the data were recorded with highly efficient triggers requiring the presence of an electron or
muon candidate with $\pt > 20 \gev$. Lower thresholds were used for the early data.

Each energy cluster reconstructed in the electromagnetic compartment of the calorimeter with $\et > 20\gev$ and $|\eta| < 2.47$
is considered as an electron candidate if it loosely matches with an inner detector track.
The electron direction is defined as that of the reconstructed track and its energy as that of the cluster.
The intrinsic resolution of the energy measurement is about 2\% at 50~\gev,
improving to approximately 1\% at 200 \gev.
Electron candidates with clusters containing cells overlapping with the
few problematic regions of the calorimeter readout are removed. This reduces the acceptance by 8\%.

Electrons are further identified based on lateral shower shapes in the first two layers
of the electromagnetic part of the calorimeter
and the fraction of energy leaking into the hadronic compartment.
A hit in the first pixel layer is also required to reduce background from photon conversions in the
inner detector material.
These requirements give about 89\% identification efficiency for electrons with $\et > 25 \gev$\
and a 1/5000 probability to falsely identify jets as electrons before isolation requirements are
imposed~\cite{atlas:wz2010}.

Muon tracks can be reconstructed independently in both the inner detector and muon spectrometer, and
the muons used in this study are required to have matching tracks in both systems.
The high-\pt\ resolution of the inner detector and muon spectrometer systems is sensitive to detector alignment.
The muons used for this analysis are restricted to those which pass through the barrel part of the muon spectrometer,
$|\eta| < 1.05$, where the muon spectrometer alignment is best understood,
in particular using high-energy cosmic rays~\cite{atlas:cosmic_muon}.
The momentum of the muon is obtained from the muon spectrometer and the average momentum resolution is currently about
20\% at $\pt = 1\tev$.
Muons are required to have hits in all three muon stations to ensure this precise measurement of the momentum.
About 80\% of the muons in the barrel are reconstructed,
with most of the loss coming from regions with limited detector coverage.

For the electron channel, the \mettext\ is obtained from a vector sum over calorimeter cells associated
with topological clusters~\cite{atlas:csc}:
\begin{linenomath}
\begin{equation}
\vecmet = \vecmetcalo = - \vecsumettopo.
\label{eqn:metcal}
\end{equation}
\end{linenomath}
In the muon channel, most of the muon energy is not deposited in the calorimeter and the
\mettext\ is obtained from
\begin{linenomath}
\begin{equation}
\vecmet = \vecmetcalo - \vecptmu + \vecetlossmu,
\label{eqn:metcalmu}
\end{equation}
\end{linenomath}
where the second term in this vector sum subtracts the muon transverse momentum and the last corrects for
the transverse component of the energy deposited in the calo\-rimeter by the muon which is included in both
of the first two terms.
The energy loss is estimated by integrating the amount of material traversed and applying a calibrated
conversion from path length to energy for each material type.

This analysis makes use of all the $\sqrt{s} = 7~\tev$ data collected in 2010 that satisfy data quality
requirements which guarantee the relevant detector systems were operating properly.
The integrated luminosity for the data used in this study is 36~\ipb\ for each channel.
The uncertainty on this estimate is 11\%~\cite{atlas:lumi}.

The \wp\ signal and
the \wzbg\ boson backgrounds
are generated with
\pythia~6.421\cite{Sjostrand:2006za} using MRST LO*~\cite{mrst} parton distribution functions (PDFs).
The \ttbar\ background is generated with \mcatnlo~3.41~\cite{mcatnlo}.
\wsl\ events are generated with \comphep~\cite{comphep} using CTEQ6L1\cite{cteq6l} PDFs 
followed by \pythia\ for parton showering and underlying event generation.
For all samples, final-state photon radiation is handled by \photos~\cite{photos} and
the propagation of particles and response of the detector are evaluated using ATLAS full detector
simulation~\cite{atlas:sim} based on \geant4~\cite{geant}.

The \pythia\ signal model used as a benchmark for \wp\ has \mbox{$V-A$} SM couplings but does not include
interference between \w\ and \wp. Decays to channels other than $e\nu$ and $\mu\nu$, including $\tau\nu$,
$ud$, $sc$ and $tb$, are included in the calculation of the \wp\ and \wstar\ widths but are not explicitly
included as signal or background.

The \wpl, \wlnu\ and \zll\ cross sections are calculated at next-to-next-to-leading order QCD (NNLO)
using
FEWZ~\cite{fewz,Gavin:2010az}
with MSTW2008 PDFs~\cite{mstw}.
For the \w\ and \z, higher-order electroweak corrections (beyond the photon radiation included in the simulation)
are calculated using \horace~\cite{horace, horaceNC}.
In the high-mass region of interest, the electroweak corrections reduce the cross sections, with the reduction
increasing with mass. For $\mt>750\gev$, the electroweak corrections
reduce the \wlnu\ cross section by 6\%.
Electroweak corrections beyond final-state radiation are not included for \wp\ because the calculation for the SM W
cannot be applied directly.
The \ttbar\ cross section is calculated at near-NNLO using the results from reference~\cite{moch_uwer}
and assuming a top-quark mass of 172.5~\gev.
The signal and most important background cross sections are listed in Table~\ref{tab:xsec}.
Cross-section uncertainties for \wpl\ and the \wzbg~\cite{atlas:wz2010} and \ttbar~\cite{atlas:ttbar2010}
backgrounds are estimated from PDF error sets,
the difference between MSTW and CTEQ PDF sets, and standard variations of renormalization and factorization scales.
The uncertainties for the LO \wsl\ cross sections include only the contributions from the PDFs.

\begin{table}[!tb]
\caption{
Calculated values of \xbr\ for \wp, \wstar\ and the leading backgrounds.
The value for $\ttbar\rightarrow\ell X$ includes all final states with at least one
lepton ($e$, $\mu$ or $\tau$).
The others are exclusive and are used for both $\ell=e$ and $\ell=\mu$.
}
\label{tab:xsec}
\begin{center}
\begin{tabular}{l|c|c|l}
\hline
\hline
        &       & Mass        &             \\
Process & Order &      [\gev] & $\xbr$ [pb] \\
\hline
\multirow{6}{*}{\wpl} & \multirow{6}{*}{NNLO}
   & \phantom{0}500 & \phantom{000}17.25   \\
 & & \phantom{0}750 & \phantom{0000}3.20   \\
 & & \phantom{}1000 & \phantom{0000}0.837  \\
 & & \phantom{}1250 & \phantom{0000}0.261  \\
 & & \phantom{}1500 & \phantom{0000}0.0887 \\
 & & \phantom{}1750 & \phantom{0000}0.0325 \\
\hline
\multirow{6}{*}{\wsl} & \multirow{6}{*}{LO}
   & \phantom{0}500 & \phantom{000}10.8    \\
 & & \phantom{0}750 & \phantom{0000}2.10   \\
 & & \phantom{}1000 & \phantom{0000}0.559  \\
 & & \phantom{}1250 & \phantom{0000}0.175  \\
 & & \phantom{}1500 & \phantom{0000}0.0595 \\
 & & \phantom{}1750 & \phantom{0000}0.0212 \\
\hline
\wlnu\ & NNLO & & \phantom{}10460 \\
\zgll\               & \multirow{2}{*}{NNLO} & & \multirow{2}{*}{\phantom{00}989} \\
($m_{\zg}>60 \gev$)  & & & \\
\multirow{2}{*}{\ttbarl}   & Near-      & & \multirow{2}{*}{\phantom{000}89.4} \\
                           &      NNLO  & &                                    \\
\hline
\hline
\end{tabular}
\end{center}
\end{table}

Except for QCD and cosmic-ray contamination, expected signal and background levels are evaluated with
simulated samples and normalized using the aforementioned cross sections and the integrated luminosity of the data.
The same reconstruction and event selection are applied to both data and simulated samples.

Events are required to have a primary vertex reconstructed from at least three tracks with
$\pt > 150$~\mev\ and longitudinal distance less than 150~mm from the center of the
collision region.
Spurious tails in \mettext\ arising from calorimeter noise and other detector problems are suppressed by checking the quality
of each reconstructed jet and discarding events where any jet has a shape indicating such problems
(following Ref.~\cite{atlas:plhc:jet_cleaning}).
Events are required to have exactly one candidate electron or one candidate muon, defined as follows.
A candidate electron is one reconstructed with $E_{T} > 25$~\gev,
$|\eta| < 1.37$ or $1.52 < |\eta| < 2.40$. 
A muon is considered a candidate if it has $\pt > 25$~\gev, $|\eta| < 1.05$ and has matching tracks in
the inner detector and muon spectrometer.
In addition, the inner detector track associated with the electron or muon is required to be compatible
with originating from the primary vertex, specifically with transverse distance of closest approach $|r_0|<1$~mm
and longitudinal distance at this point $|z_0|<5$~mm.

The above requirements constitute the event preselection criteria.
To suppress the QCD background, the lepton is required to be isolated. In the electron channel,
the isolation energy is measured with the calorimeter in a cone $\Delta R < 0.4$
($\Delta R = \sqrt{(\Delta \eta)^2 + (\Delta \varphi)^2}$) around the electron track
and the requirement is \mbox{$\sumet < 10 \gev$,}
where the sum excludes the core energy deposited by the electron and is  
corrected to account for leakage of the electron energy outside this core.
In the muon channel, the isolation energy is measured using inner detector tracks with $\pttrack > 1$~\gev\ in a cone
$\Delta R < 0.3$ around the muon track.
The isolation requirement is
\mbox{$\sumpttrack < 0.05~\pt$},
where the muon track is excluded from the sum.
The scaling of the threshold with the muon \pt\ 
reduces efficiency losses due to radiation from the muon at high \pt.

Finally, a \mettext\ threshold is applied to further suppress the QCD background.
In both channels, a fixed threshold is applied:
\mbox{$\met > 25 \gev$}.
In the electron channel, where QCD jets may be misidentified as electrons,
a scaled threshold is also applied:
\mbox{$\met > 0.6~\et$}.
Taken together, all the above constitute the final selection requirements.

\begin{figure*}[]
  \centering
  \includegraphics[width=0.48\textwidth]{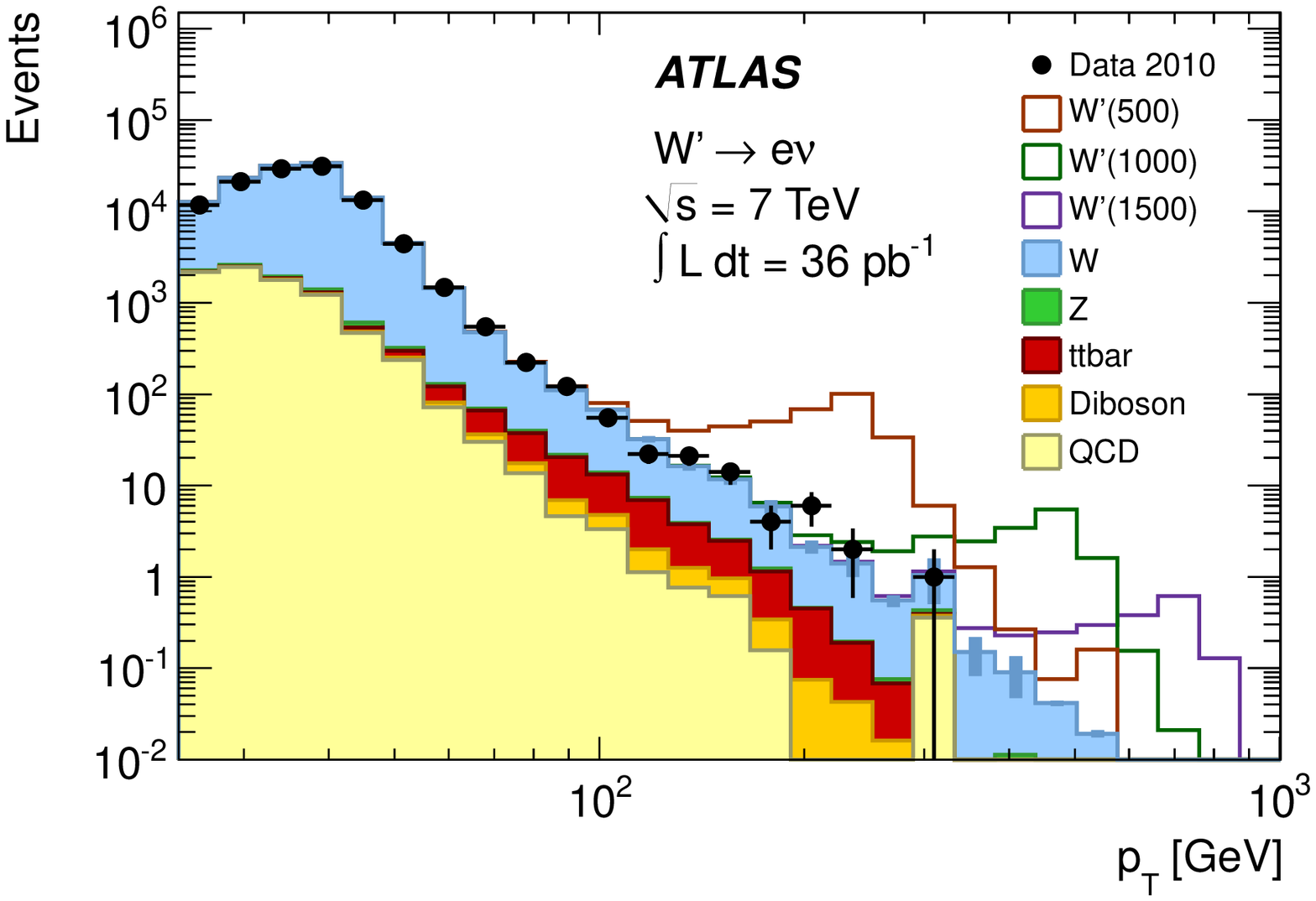}
  \includegraphics[width=0.48\textwidth]{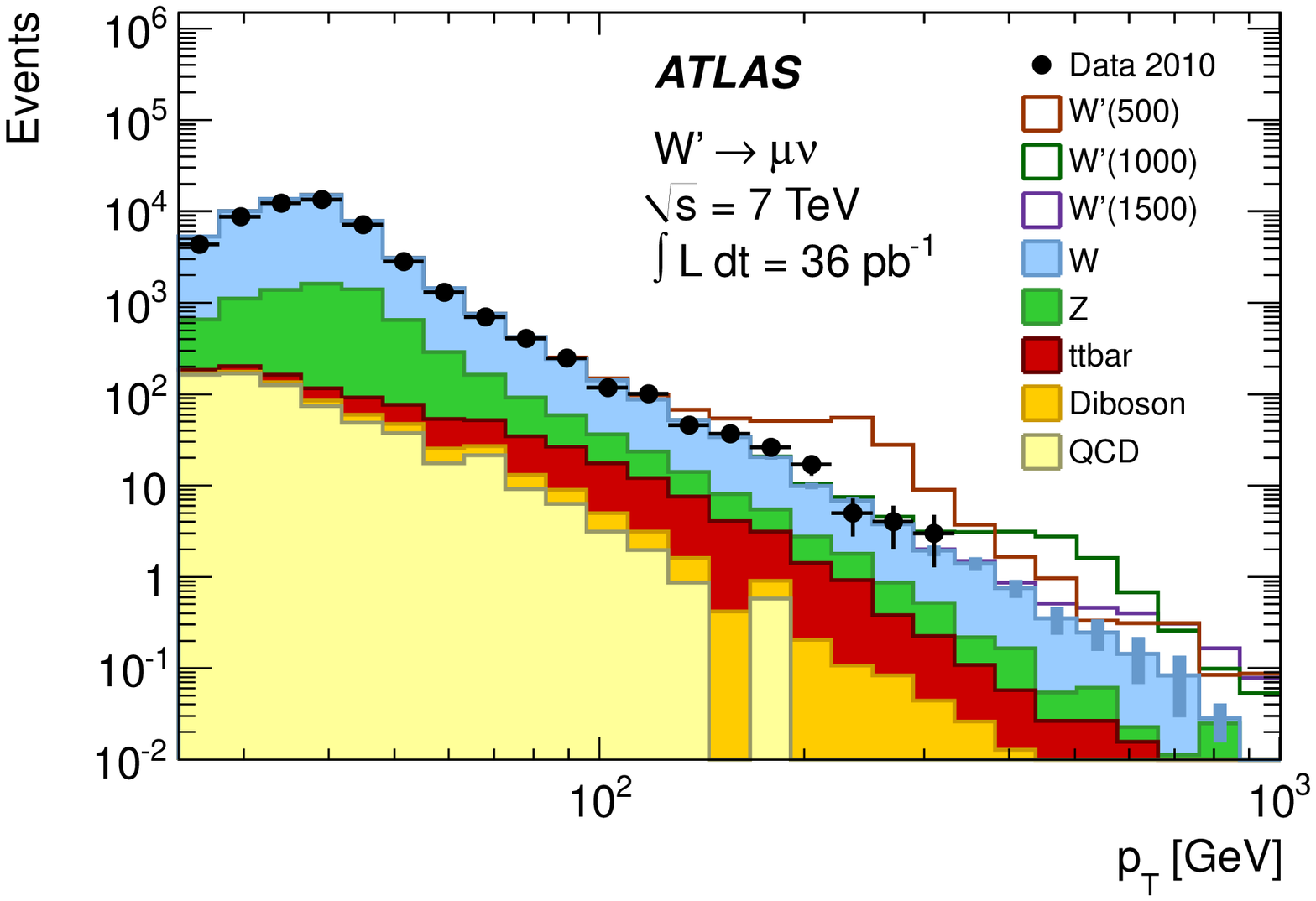}
  \includegraphics[width=0.48\textwidth]{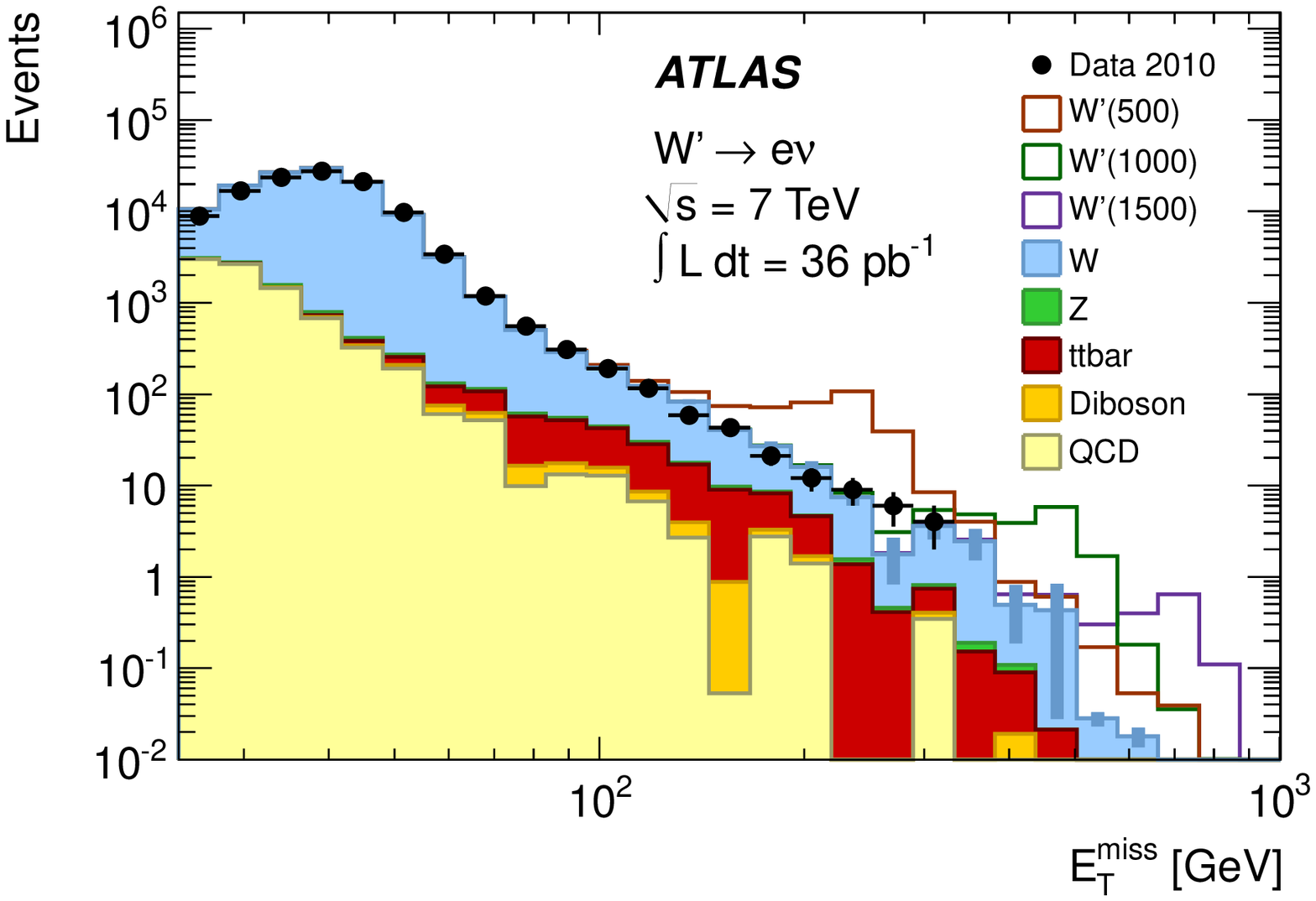}
  \includegraphics[width=0.48\textwidth]{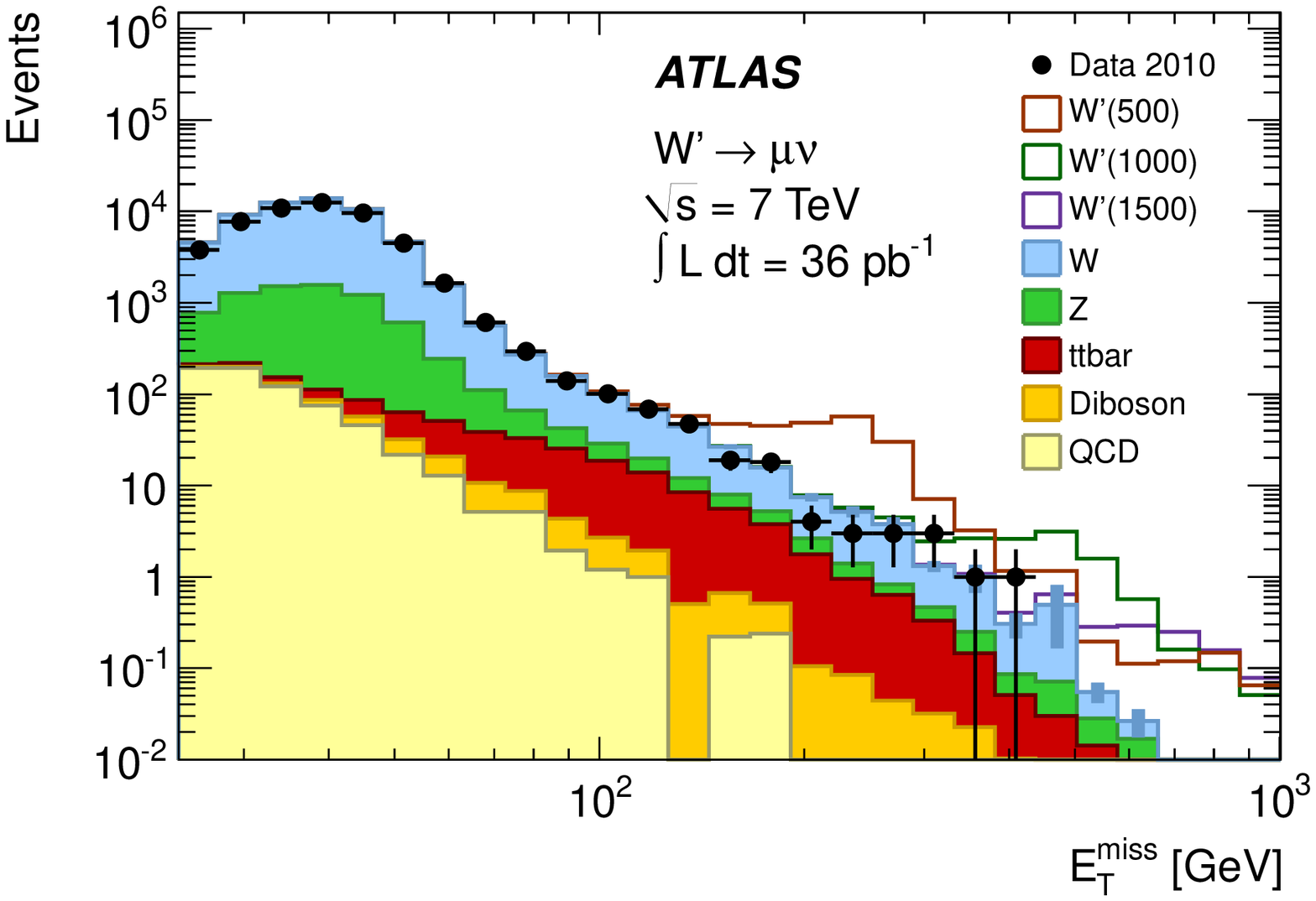}
  \includegraphics[width=0.48\textwidth]{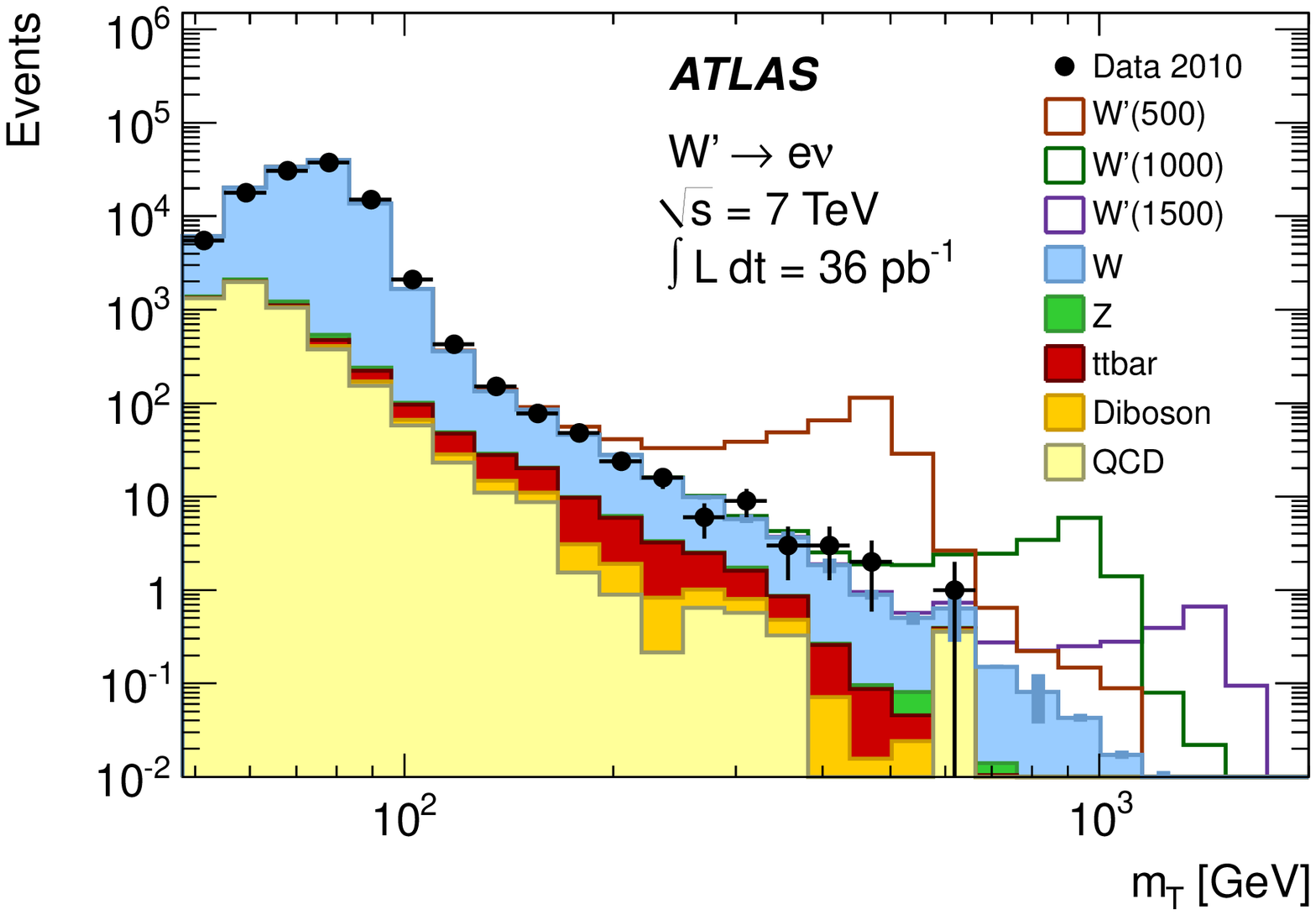}
  \includegraphics[width=0.48\textwidth]{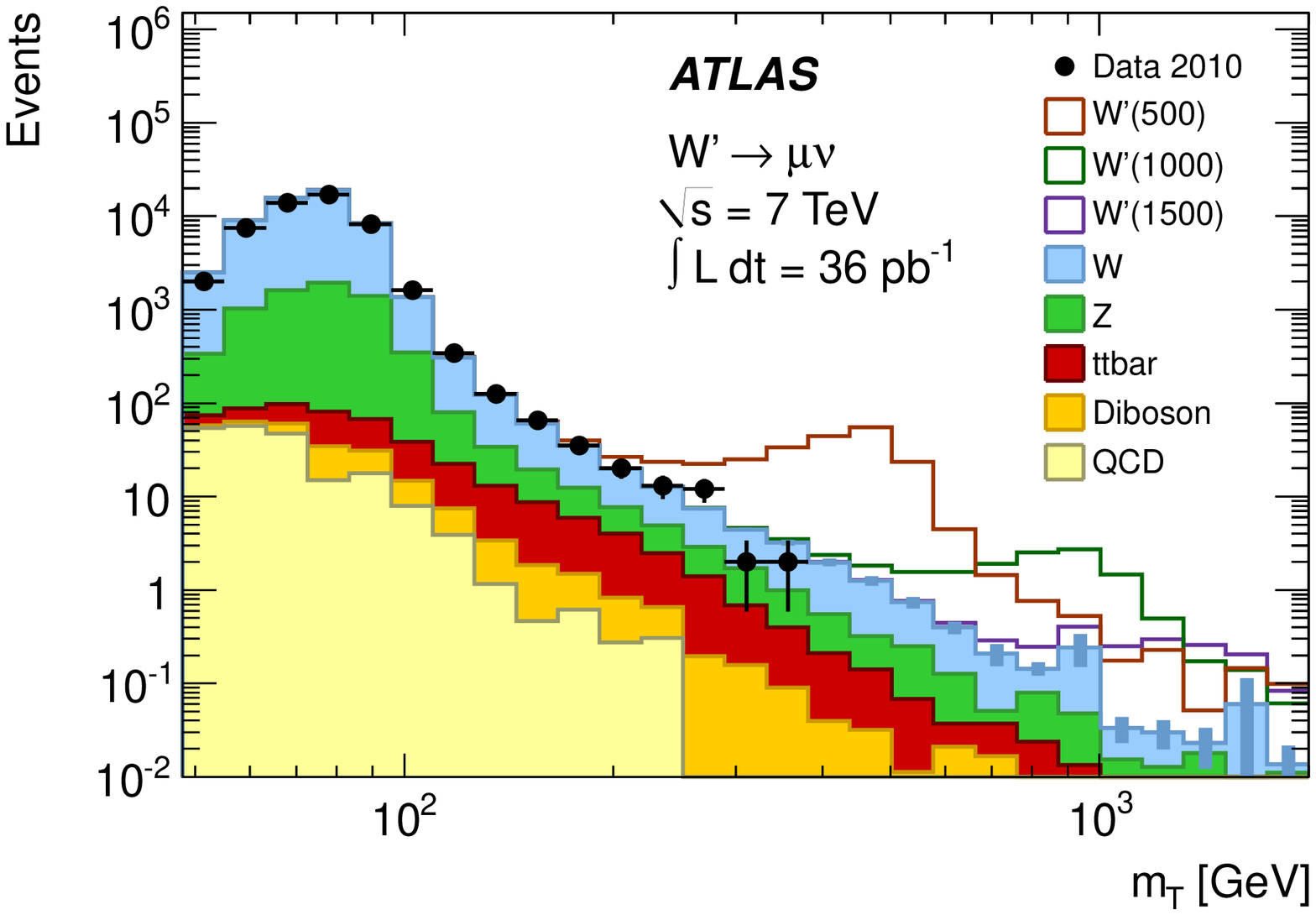}
  \caption{Spectra of \pt\ (top), \mettext\ (center) and \mt\ (bottom)
  for the electron (left) and muon (right) channels after final event selection.
  The points represent ATLAS data and the filled histograms show the stacked backgrounds.
  Both direct production of leptons and indirect from \tauleps\ are included.
  Open histograms are \wp\ signals added to the background with masses in \gev\ indicated in parentheses in the legend. 
  The QCD background is estimated from data.
  The signal and other background samples are normalized using the integrated luminosity of the data
  and the NNLO (near-NNLO for \ttbar) cross sections listed in Table~\ref{tab:xsec}.
  \label{fig:final_mt}
  }
\end{figure*}

Figure~\ref{fig:final_mt} shows the \pt, missing \et, and \mt\ spectra in both channels after final selection
for the data, for the expected background, and for three examples of \wp\ signals at different masses.
The agreement between the data and expected background is good.
Table~\ref{tab:nbg} shows as an example how different sources contribute to the background
for $\mt>750 \gev$, which is the region used to search for a \wp\ or \wstar\ with a mass of 1500~\gev.
There are significant differences between the background levels in the electron and muon channels. The
background from \wlnu\ and \ttbar\ is higher in the muon channel because of the worse momentum resolution for high-\pt\ muons.
The difference is even larger for the \zll\ background because there is additionally a much larger chance that
one lepton is lost due to the restricted acceptance in \eta. The QCD background in the electron channel is
less than that in the muon channel because of the tighter electron selection criteria: an isolation threshold
that is not scaled with \pt\ and the addition of a scaled \mettext\ threshold.

\begin{table}[!bt]
\caption{
Expected number of events from the various background
sources in both decay channels
for \mbox{$\mt>750 \gev$}, i.e. for \wps\ with a mass of 1500~\gev.
The \wlnu\ and \zll\ entries include the expected contributions from the $\tau$-lepton.
The uncertainties are statistical.
}
\label{tab:nbg}
\begin{center}
\begin{tabular}{l | l|l}
\hline
\hline
                & \multicolumn{1}{c|}{$e\nu$} & \multicolumn{1}{c}{$\mu\nu$} \\
\hline
\tspace
\wlnu           & 0.145\phantom{0} $\pm$ 0.001   & 0.43\phantom{0}  $\pm$ 0.10 \\
\zll            & 0.0001\phantom{} $\pm$ 0.0001  & 0.11\phantom{0}  $\pm$ 0.02 \\
diboson         & 0.011\phantom{0} $\pm$ 0.001   & 0.01\phantom{0}  $\pm$ 0.01 \\
\ttbar          & 0.003\phantom{0} $\pm$ 0.003   & 0.05\phantom{0}  $\pm$ 0.02 \\
QCD             & $0.001^{~~+0.004}_{~~-0.001}$  & $0.02\phantom{0} ^{~+0.05}_{~-0.01}$ \\
Cosmic ray      &                                & 0.006\phantom{} $\pm$ 0.003 \\
\hline
\tspace
Total           & 0.159\phantom{0} $\pm$ 0.005   & 0.62\phantom{0} $\pm$ 0.11 \\
\hline
\hline
\end{tabular}
\end{center}
\end{table}

\begin{figure}[]
  \centering
  \includegraphics[width=0.48\textwidth]{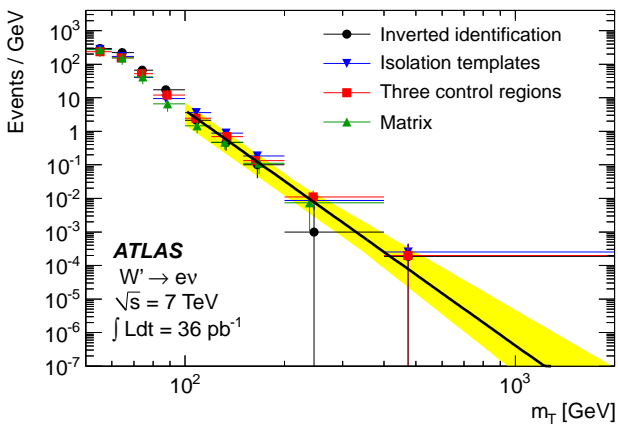}
  \caption{
  Estimated QCD background as a function of \mt\ in the electron channel after final selection
  as obtained from the four data-driven methods (see text).
  The power-law fit to all four sets of results and its $1\sigma$ uncertainty band are also shown.
  \label{fig:el_qcd}
  }
\end{figure}

In the electron channel, four techniques are used to estimate the QCD
background level from data
through the use of subsidiary samples which are disjoint from the
analysis region.
In the ``Inverted identification'' technique, the distributions of
the QCD background as a function of \pt, \mettext,
or \mt\ are estimated from events which pass relaxed
identification criteria but fail the normal selection.
The normalization is obtained by fitting the \mettext\ distribution plus
the estimates for EW and \ttbar\ to the observed data.
The other techniques are described elsewhere: ``Isolation templates"~\cite{atlas:wz2010},
``Three control regions"~\cite{atlas:photon2010},
``Matrix"~\cite{atlas:electron_performance, atlas:ttbar2010}.
Figure~\ref{fig:el_qcd} shows the estimates obtained from all four techniques after final selection
as a function of \mt\ along with the power-law fit to all four sets of results and its $1\sigma$ uncertainty band.
The extrapolation of this fit and uncertainty band provides the estimate of the QCD background level
and uncertainty in the high-\mt\ region used for the limit calculations.

\begin{figure}[]
  \centering
  \includegraphics[width=0.48\textwidth]{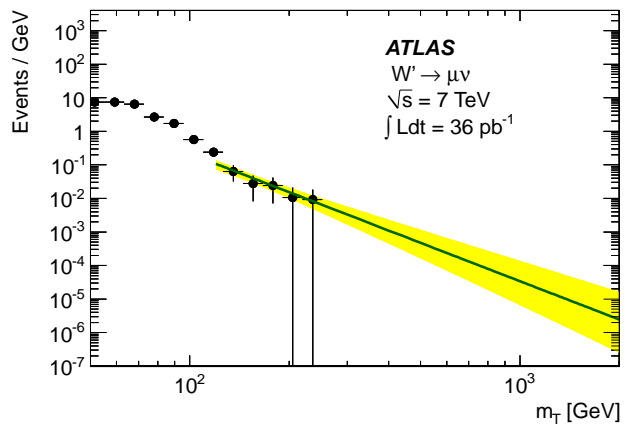}
  \caption{
  Estimated QCD background as a function of \mt\ in the muon channel after final selection
  as obtained from the data-driven method (see text).
  The unbinned power-law fit to the data and its $1\sigma$ uncertainty band are also shown.
  \label{fig:mu_qcd}
  }
\end{figure}

The shape of the QCD background for the muon channel is evaluated by starting with the muon
preselection and replacing the isolation threshold
with a range of values in the non-isolated region: $0.2<\sumpttrack/\pt<0.4$.
The normalization of the QCD background is determined by fitting
the resulting \mettext\ spectrum plus the EW and \ttbar\ predictions from simulation to the data after
final selection, excluding the \mettext\ threshold.
The isolation range used to determine the shape is varied to determine the uncertainty in the prediction
for the QCD background level.
Figure~\ref{fig:mu_qcd} shows the predicted background level after final selection as a function of \mt\
along with the unbinned power-law fit and its $1\sigma$ uncertainty band.
The range of \mt\ used for the fit is the one which gives largest values for the upper end of this band.
The lower end of the uncertainty band corresponds to a negligible background level for all fits.
The extrapolation of the fit and uncertainty band provides the QCD background level and uncertainty 
in the high-\mt\ region used for the limit calculations.

Cosmic rays can mimic the signal in the muon channel if the muon is only reconstructed on one side of the detector.
Most of this background is rejected by the requirement that the muon pass close to the primary vertex and the remainder is
estimated by looking at the rate away from the vertex.
The measured rate after final selection is less than 2\% of the total background
for any \mt\ threshold relevant to this analysis.


The data show no evidence for any excess above SM expectations and are used to set limits on \xbr\ for
\wp\ and \wstar\ production with the masses listed in Table~\ref{tab:xsec}.
The limits are evaluated using a single-bin likelihood analysis, i.e. by counting events with $\mt > 0.5~\mwps$.
The expected number of events in each channel is
\begin{linenomath}
\begin{equation}
\nexp = \effsig \lint \xbr + \nbg,
\end{equation}
\end{linenomath}
where \lint\ is the integrated luminosity of the data sample and \effsig\ is the event selection efficiency,
i.e. the fraction of events that pass final event selection criteria and have \mt\ above threshold.
\nbg\ is the expected number of background events.
Using Poisson statistics, the likelihood to observe \nobs\ events is:
\begin{linenomath}
\begin{equation}
  {\cal L}(\xbr) = \frac{(\lint \effsig \xbr + \nbg)^{\nobs} e^{-(\lint \effsig \xbr + \nbg)}}{\nobs!}
\label{eqn:lhood}
\end{equation}
\end{linenomath}
and this expression is used to set limits on \xbr.
Uncertainties are handled
by introducing nuisance parameters and multiplying by the
probability density function (pdf) characterizing that uncertainty:
\begin{linenomath}
\begin{equation}
  {\cal L}(\xbr,\nps) =   {\cal L}(\xbr)   \prod \pdfi,
\label{eqn:lhoodn}
\end{equation}
\end{linenomath}
where \pdfi\ is the Gaussian pdf for nuisance parameter \npi.
The nuisance parameters are taken to be the explicit dependencies: \lint, \effsig\ and \nbg.
Correlations between these are neglected. This is justified by the small effect that the
nuisance parameters themselves have on the limits, as demonstrated below.

The fraction of fully simulated signal events that pass final selection and are above \mt\ threshold provides an initial
estimate of the expected numbers of events for each mass.
Small corrections are made to account for differences between the kinematical distributions at NNLO
(obtained from FEWZ) and those in the LO simulation.
The largest correction is around 4\%.
Contributions from \mbox{\wptau} with the \taulep\ decaying leptonically have been neglected and would increase the \wp\
selection efficiencies by 3-4\%.

The EW and \ttbar\ background predictions are also obtained from full simulation, normalized to the integrated
luminosity of the data.
For the EW background, small corrections are again made to account for differences between kinematical distributions
in LO simulation and higher order calculations, now using NLO MCFM~\cite{mcfm} because the present version of
FEWZ does not provide reliable values far from the resonance peak.
The background level for each mass is obtained by adding the small QCD
and cosmic-ray contributions to these values.

The uncertainties on \effsig\ and \nbg\ account for experimental and theoretical systematic effects as well as
the statistics of the simulation samples.
The experimental systematic uncertainties include efficiencies for lepton trigger, reconstruction,
impact parameter and isolation as well as event vertex reconstruction.
Lepton momentum and \mettext\ response, characterized by scale and resolution, are also included.
Most of these performance metrics are measured at relatively low \pt\ and their values are extrapolated to
the high-\pt\ regime relevant to this analysis.
The uncertainties due to these extrapolations are included but are too small to significantly affect the \wps\ limits.
The uncertainties on the QCD and cosmic-ray background estimates also contribute to \nbg.
Theoretical systematic uncertainties arise from the calculation of cross sections and their kinematical
distributions,
lepton isolation, and the distribution of the ratio of neutrino to lepton \pt\ which affects the scaled
\mettext\ selection efficiency.

Table~\ref{tab:syst_summary} summarizes the uncertainties on the event-selection efficiencies 
and background levels for a \wp\ signal with $\mwp=1500 \gev$ (i.e. for $\mt>750$~GeV).

\begin{table*}[]
\caption{
  Relative uncertainties on the event-selection efficiency and background level for 
  a \wp\ with a mass of 1500~GeV.
  The most important uncertainties are indicated in bold.
  The last row gives the total uncertainties.
\label{tab:syst_summary}
}
\begin{center}
\begin{tabular}{l|rr|rr}
\hline
\hline
 & \multicolumn{2}{c|}{\effsig} & \multicolumn{2}{c}{\nbg} \\
 Source                      &  \multicolumn{1}{c}{$e\nu$}  & \multicolumn{1}{c|}{$\mu\nu$} &  \multicolumn{1}{c}{$e\nu$}  & \multicolumn{1}{c}{$\mu\nu$} \\
\hline
 \mettextcap\ scale          &       0.1\%  &       0.1\%  &       1.1\%  &  {\bf 3.4\%} \\
 Trigger efficiency          & {\bf  1.0\%} &  {\bf 0.7\%} &       1.0\%  &       0.7\%  \\
 Reco. and id. efficiency    & {\bf  3.6\%} &  {\bf 1.6\%} &  {\bf 3.6\%} &       1.3\%  \\
 Isolation leakage           & {\bf  2.7\%} &              &  {\bf 3.4\%} &              \\
 Energy/momentum resolution  &       0.1\%  &       0.4\%  &       2.4\%  &  {\bf 3.1\%} \\
 Energy/momentum scale       & {\bf  0.8\%} &       0.1\%  &  {\bf 6.6\%} &       0.1\%  \\
 Correlated misalignment     &              &       0.6\%  &              &  {\bf 3.3\%} \\
 QCD background              &              &              &       2.2\%  &  {\bf 7.7\%} \\
 Monte Carlo statistics      & {\bf  1.7\%} &  {\bf 1.6\%} &       2.2\%  & {\bf 16.6\%} \\
 Cross section (shape/level) & {\bf  0.7\%} &  {\bf 0.7\%} &  {\bf 8.5\%} &  {\bf 7.7\%} \\
 Isolation                   & {\bf  1.5\%} &  {\bf 1.5\%} &       1.0\%  &       1.0\%  \\
 Other                       &       0.2\%  &       0.4\%  &       0.4\%  &       0.9\%  \\
\hline
 All                         &       5.3\%  &       3.0\%  &      12.6\%  &      20.7\%  \\
\hline
\hline
\end{tabular}
\end{center}
\end{table*}

For \effsig, most of the uncertainty in the electron channel comes from electron identification
except for the higher masses where the isolation leakage is also important. The total is
less than 6\% for all \wps\ masses and has a negligible effect on the limit evaluation.
The signal uncertainties are even smaller in the muon channel.
For \nbg, the dominant uncertainties in the electron channel come from the electron energy scale and the
cross-section calculation.
For the muon channel, the simulation statistics followed by the uncertainties on the QCD background
and cross-section calculation dominate.
The first is large because momentum smearing pushes events with low \mt, and hence higher cross section,
into the high-\mt\ bins used in the limit evaluation.
The cross-section uncertainties are large (around 8\% in Table~\ref{tab:syst_summary})
because it is the high-mass tail that is relevant to this analysis.

Limits for 95\% CL (confidence level) exclusion on \xbr\ for each \wp\ and \wstar\ mass and decay channel are set using the
likelihood function in Eq.~\ref{eqn:lhoodn} as input to the estimator
$\cls = \clsb/\clb$~\cite{junk}.
The inputs for the limit calculation are \lint, \effsig, \nbg, \nobs\ and the uncertainties on
the first three.
Except for \lint\ and its uncertainty, these inputs are all listed in
Table~\ref{tab:limit_input}.
The table also lists the predicted numbers of signal events, \nsig, with their uncertainty including both that of \effsig\ and
the cross-section calculation.
The uncertainties on \effsig, \nbg\ and \nsig\ account for all relevant experimental and theoretical effects except
for integrated luminosity which is included separately to allow for the correlation between signal and background.
The numbers of observed events are in good agreement with the expected numbers of background events for all
mass bins in the electron channel and for the lowest bin in the muon channel. A discrepancy is observed in the muon channel for
$\mt > 750 \gev$ where 5.48 muon events are predicted and none are observed, a result for which the Poisson probability
is only 0.4\%.
However, the muon \pt\ spectrum in Fig.~\ref{fig:final_mt} shows no evidence of any discrepancy between data
and predicted background at high \pt, confirming that, as expected, the muon efficiency remains stable at high \pt.


\begin{table*}[!htbp]
\caption{Inputs for the \wpsl\ \xbr\ limit calculations for an integrated luminosity of 36 \ipb.
The first two columns are the \wps\ mass and decay mode.
The next four are the corrected signal selection efficiency, \effsig, and the prediction
for the number of signal events, \nsig, obtained with this efficiency.
The last two columns are the expected number of background
events, \nbg, and the number of events observed in data, \nobs.
The uncertainties for \nsig\ and \nbg\ include contributions from the uncertainties in the
cross sections but not from the integrated luminosity.
\label{tab:limit_input}
}
\begin{center}
\begin{tabular}{rr| @{~~}r@{ $\pm$ }r | @~r@{ $\pm$ }r |
                    @{~~}r@{ $\pm$ }r | @~r@{ $\pm$ }r |
                    @{~~}r@{ $\pm$ }r | r}
\hline
\hline
$m$      &           & \multicolumn{4}{c}{\effsig} & \multicolumn{4}{c}{\nsig} \\
~[\gev]  & decay     & \multicolumn{2}{c}{\wp} & \multicolumn{2}{c}{\wstar}
                     & \multicolumn{2}{c}{\wp} & \multicolumn{2}{c}{\wstar}
                     & \multicolumn{2}{c}{\nbg} & \nobs \\
\hline
\hline
\multirow{2}{*}{500}
         & $e\nu$    & 0.556 & 0.024 & 0.530 & 0.022 & 349\phantom{.00} & 30\phantom{.00}  & 208\phantom{.00} & 18\phantom{.00}  & 21.5\phantom{00} & 2.0\phantom{00}  & 24 \\
         & $\mu\nu$  & 0.339 & 0.008 & 0.265 & 0.005 & 212\phantom{.00} & 17\phantom{.00}  & 104\phantom{.00} &  8\phantom{.00}  & 20.3\phantom{00} & 1.1\phantom{00}  & 16 \\
\hline
\multirow{2}{*}{750}
         & $e\nu$    & 0.565 & 0.025 & 0.520 & 0.022 &  65.8\phantom{0} &  4.8\phantom{0}  &  39.6\phantom{0} &  3.5\phantom{0}  &  4.05\phantom{0} & 0.35\phantom{0}  &  6 \\
         & $\mu\nu$  & 0.362 & 0.009 & 0.257 & 0.005 &  42.1\phantom{0} &  2.7\phantom{0}  &  19.6\phantom{0} &  1.5\phantom{0}  &  5.48\phantom{0} & 0.44\phantom{0}  &  0 \\
\hline
\multirow{2}{*}{1000}
         & $e\nu$    & 0.562 & 0.025 & 0.516 & 0.022 &  17.1\phantom{0} &  1.4\phantom{0}  &  10.5\phantom{0} &  1.0\phantom{0}  &  1.11\phantom{0} & 0.11\phantom{0}  &  1 \\
         & $\mu\nu$  & 0.381 & 0.010 & 0.264 & 0.006 &  11.6\phantom{0} &  0.9\phantom{0}  &   5.4\phantom{0} &  0.5\phantom{0}  &  2.05\phantom{0} & 0.25\phantom{0}  &  0 \\
\hline
\multirow{2}{*}{1250}
         & $e\nu$    & 0.552 & 0.026 & 0.505 & 0.023 &   5.23\phantom{} &  0.51\phantom{}  &   3.22\phantom{} &  0.42\phantom{}  &  0.400\phantom{} & 0.054\phantom{}  &  0 \\
         & $\mu\nu$  & 0.386 & 0.011 & 0.255 & 0.006 &   3.66\phantom{} &  0.33\phantom{}  &   1.63\phantom{} &  0.20\phantom{}  &  1.01\phantom{0} & 0.17\phantom{0}  &  0 \\
\hline
\multirow{2}{*}{1500}
         & $e\nu$    & 0.530 & 0.028 & 0.488 & 0.025 &   1.71\phantom{} &  0.21\phantom{}  &   1.06\phantom{} &  0.17\phantom{}  &  0.159\phantom{} & 0.020\phantom{}  &  0 \\
         & $\mu\nu$  & 0.383 & 0.012 & 0.252 & 0.006 &   1.24\phantom{} &  0.14\phantom{}  &   0.54\phantom{} &  0.08\phantom{}  &  0.62\phantom{0} & 0.13\phantom{0}  &  0 \\
\hline
\multirow{2}{*}{1750}
         & $e\nu$    & 0.503 & 0.027 & 0.482 & 0.028 &   0.59\phantom{} &  0.09\phantom{}  &   0.37\phantom{} &  0.07\phantom{}  &  0.069\phantom{} & 0.009\phantom{}  &  0 \\
         & $\mu\nu$  & 0.360 & 0.012 & 0.254 & 0.007 &   0.43\phantom{} &  0.06\phantom{}  &   0.20\phantom{} &  0.04\phantom{}  &  0.47\phantom{0} & 0.09\phantom{0}  &  0 \\
\hline
\hline
\end{tabular}
\end{center}
\end{table*}

\begin{figure*}[!htbp]
  \centering
  \includegraphics[width=0.49\textwidth]{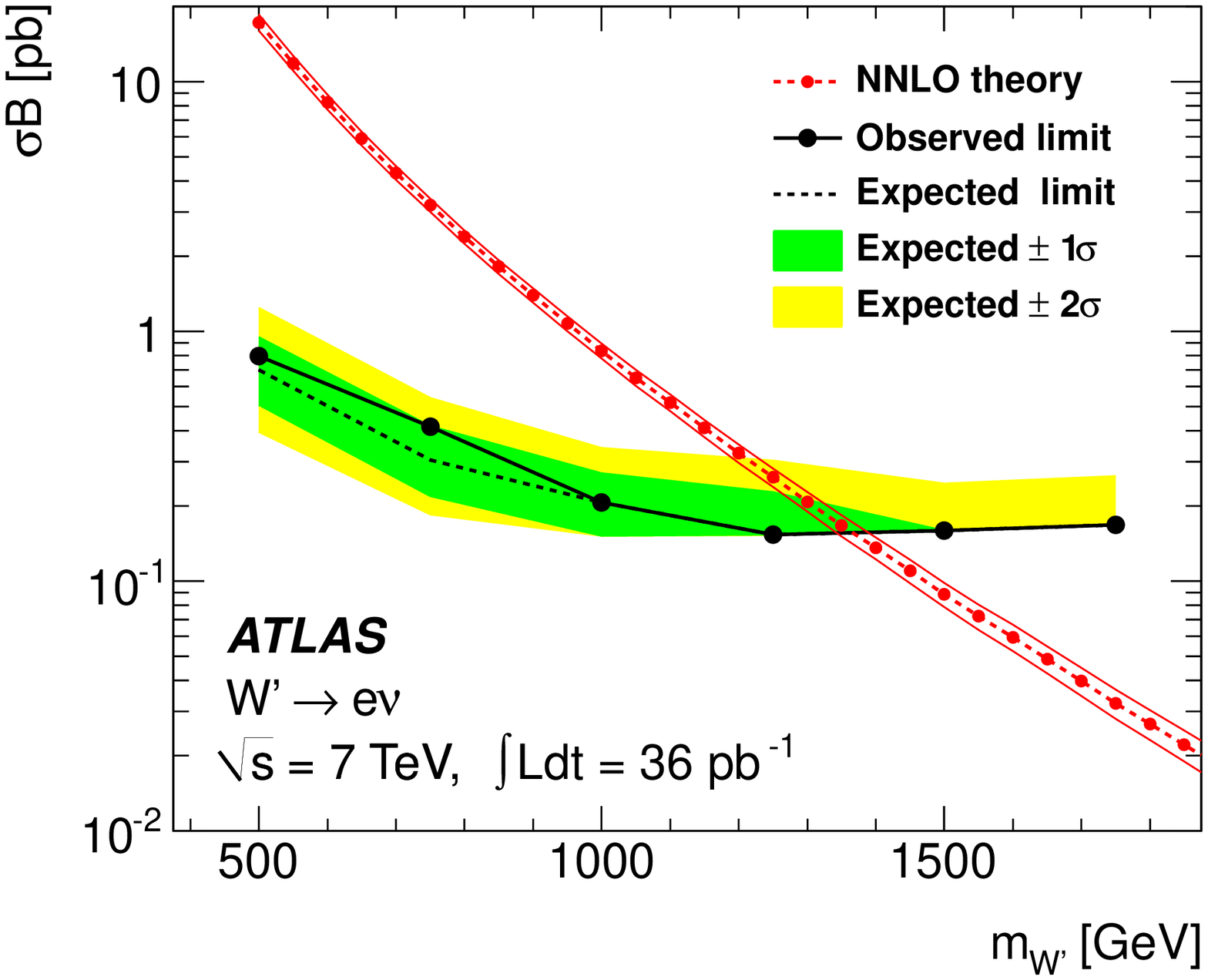}
  \includegraphics[width=0.49\textwidth]{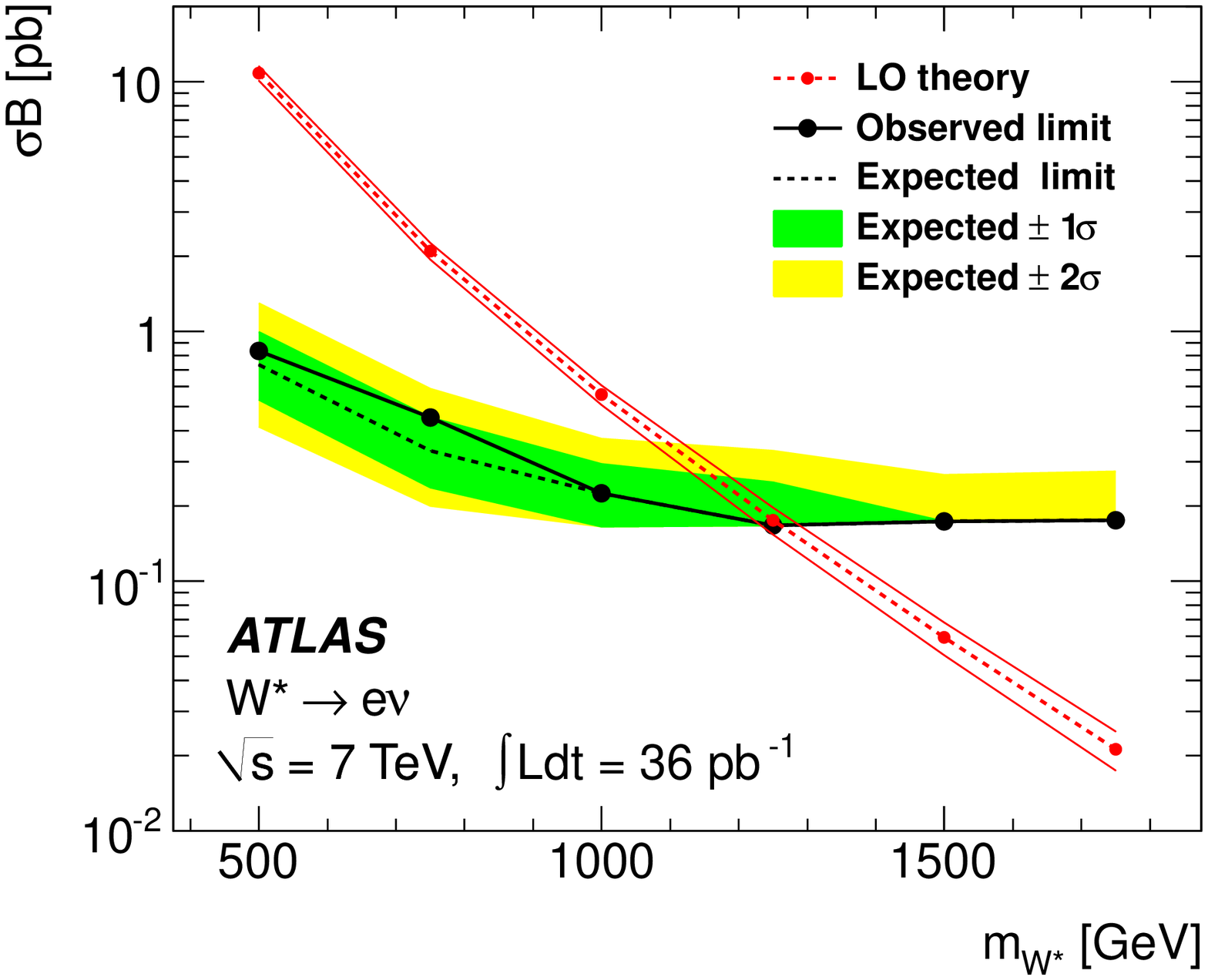}
  \includegraphics[width=0.49\textwidth]{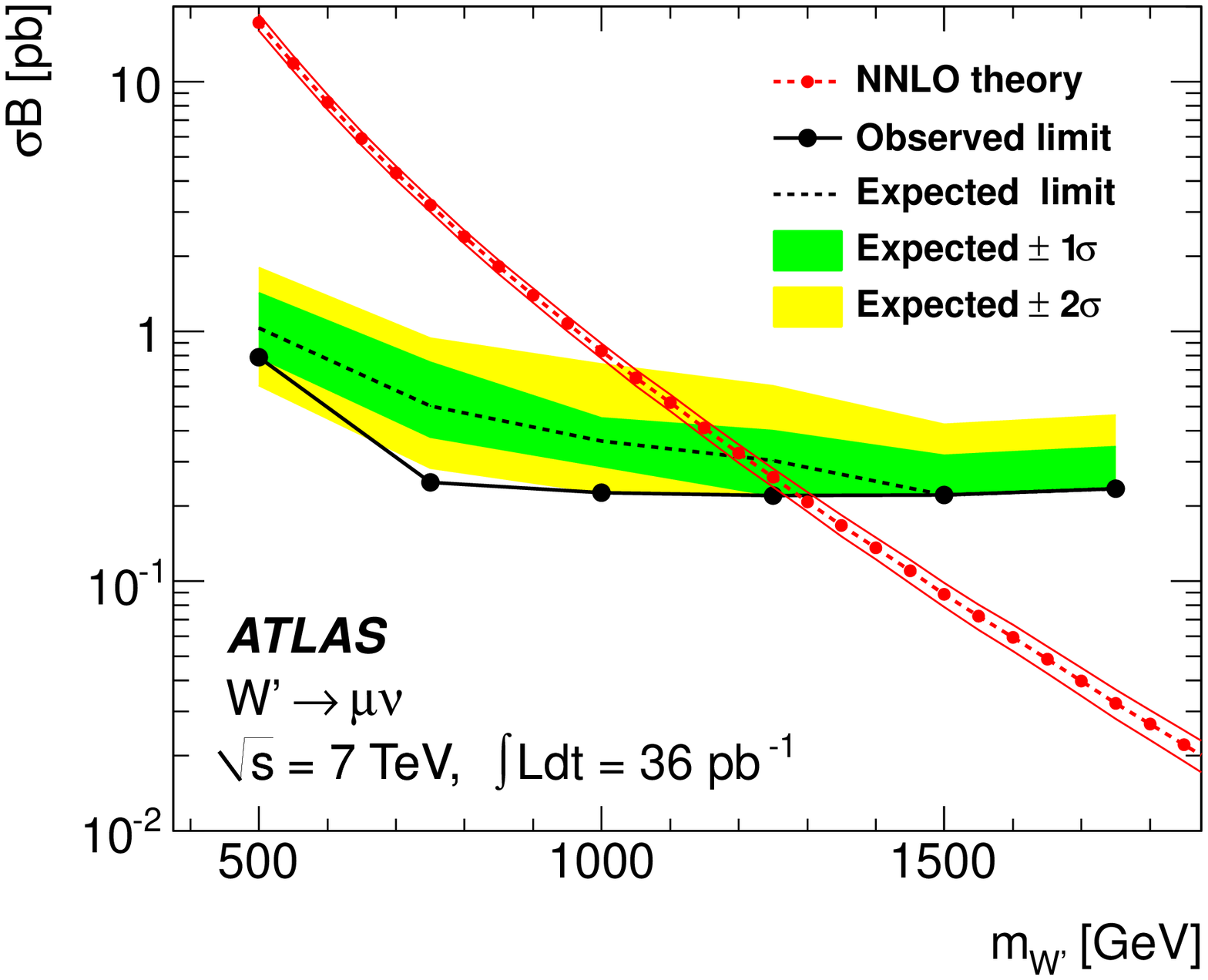}
  \includegraphics[width=0.49\textwidth]{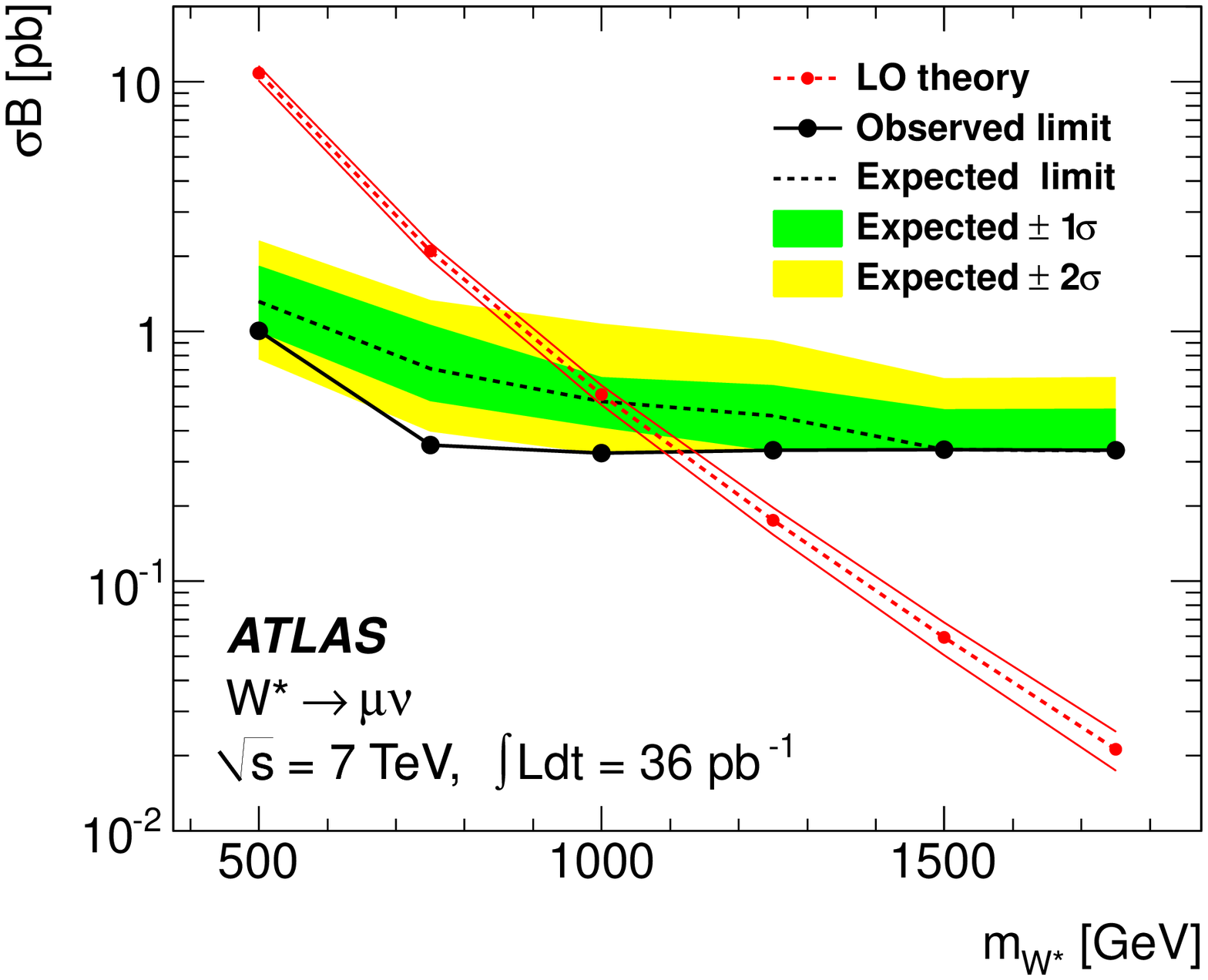}
  \includegraphics[width=0.49\textwidth]{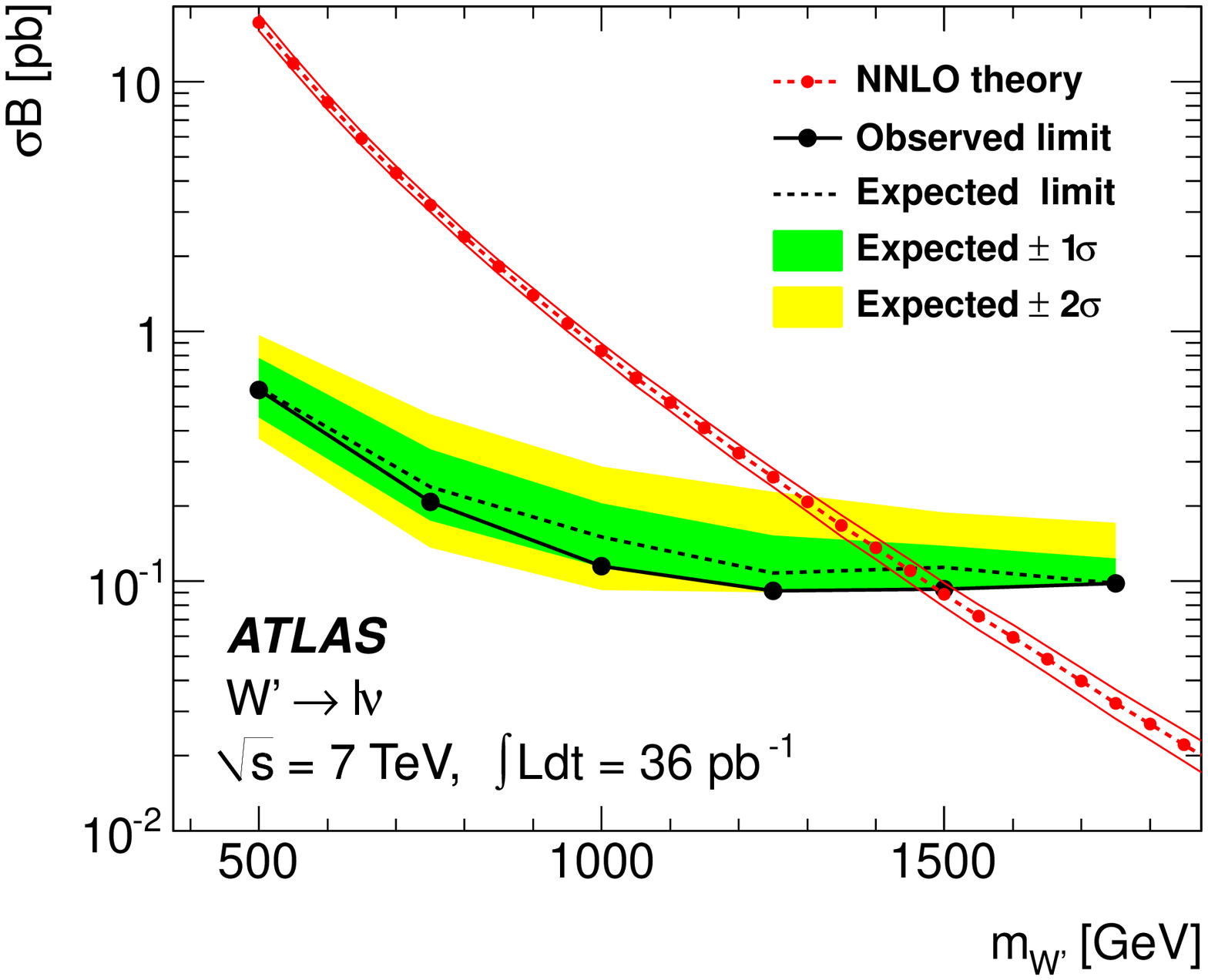}
  \includegraphics[width=0.49\textwidth]{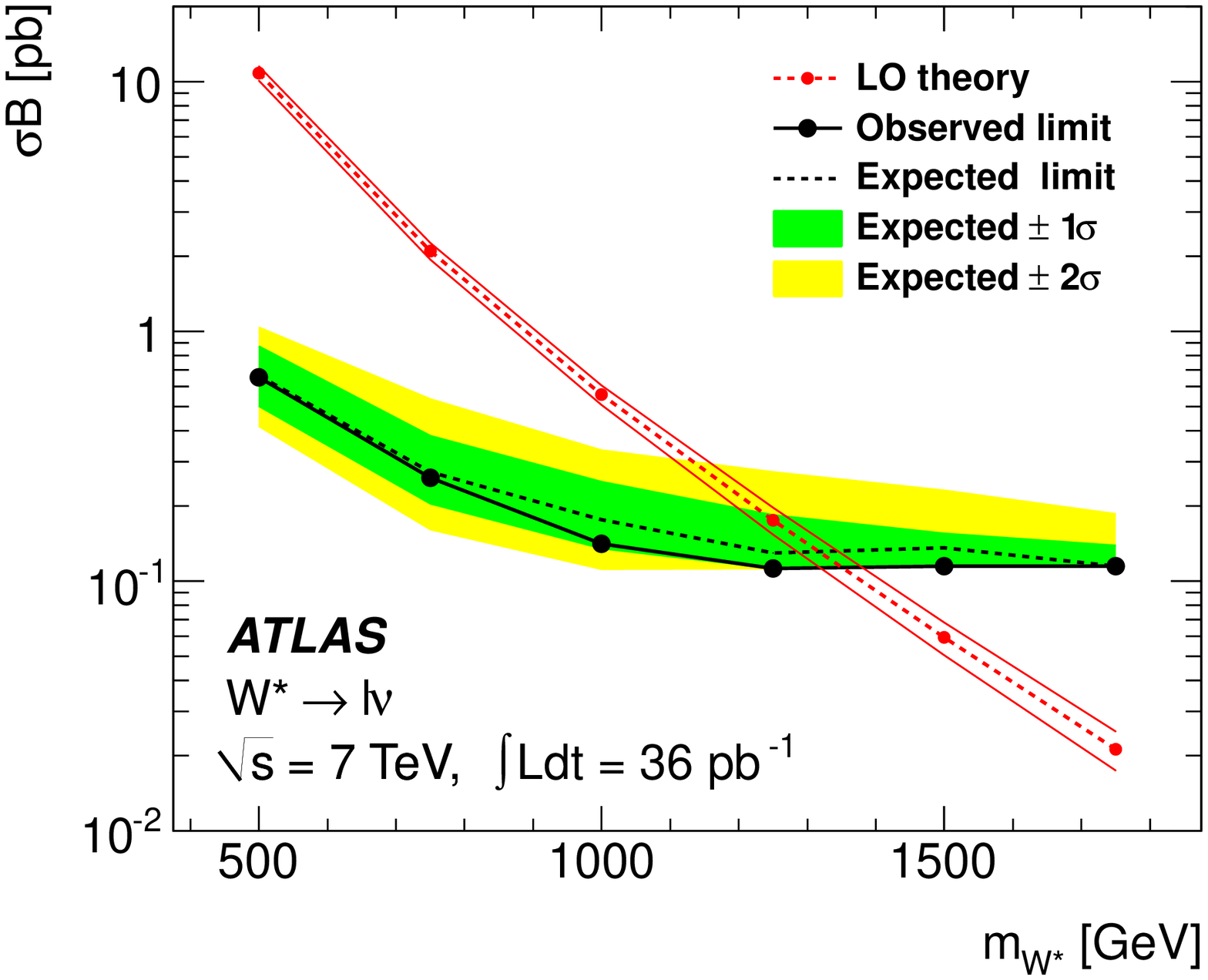}
  \caption{
Limits at 95\% CL for \wp\ (left) and \wstar\ (right) production in the
decay channels \wpse\ (top), \wpsmu\ (center), and the combination of these (bottom).
The solid lines show the observed limits with all uncertainties.
The expected limit is indicated with dashed lines surrounded by
$1\sigma$ and $2\sigma$ shaded bands.
Dashed lines show the theory predictions
(NNLO for \wp, LO for \wstar)
between solid lines indicating their uncertainties.
The \wp\ \xbr\ uncertainties are obtained by varying renormalization and factorization scales and by
varying PDFs. Only the latter are included for \wstar.
  \label{fig:limits_xbr}
  }
\end{figure*}

Table~\ref{tab:limits_xbr} and Fig.~\ref{fig:limits_xbr} show the limits obtained from these values.
The figure also shows the expected limits and the theoretical \wps\ \xbr\ as a
function of \mt\ for both channels and their combination.
The intersection between the central theoretical prediction and the observed limits
provides the 95\% CL lower limit on the mass.
Table~\ref{tab:limits_mass} presents the \wp\ and \wstar\ expected and observed mass limits for the electron and muon
decay channels and for the combination of both channels.
These limits increase by 5-10~\gev\ if the uncertainties on \effsig, \nbg\ and \lint\ are neglected.
For both channels, the effect of the \effsig\ and \nbg\ uncertainties on the limits is
small for the lowest-\mt\ bin and negligible for the others.

\begin{table}[!htbp]
\caption{Upper limits on \wp\ and \wstar\ \xbr.
         The first two columns are the mass and decay channel and the following are
         the 95\% CL limits with headers indicating the nuisance parameters
         for which uncertainties are included: S for the event selection efficiency (\effsig),
         B for the background level (\nbg), and L for the integrated luminosity (\lint).
         Columns labeled SBL include all uncertainties and are used to evaluate mass limits.
         Results are given for the electron and muon channels and the combination of the two.
\label{tab:limits_xbr}
}
\begin{center}
\begin{tabular}{rc|rrrr|rr}
\hline
\hline
         & & \multicolumn{6}{c}{95\% CL limit on \xbr\ [fb]} \\
 mass    & & \multicolumn{4}{c|}{\wp} & \multicolumn{2}{c}{\wstar} \\
 ~[\gev] &              & none & S & SB  & SBL & none & SBL \\      
\hline
\hline
         & $e\nu$    & 647 & 649 & 682  & 795   & 679 & 834 \\
 500     & $\mu\nu$  & 625 & 625 & 640  & 786   & 799 &1005 \\
         & both      & 413 & 416 & 444  & 583   & 473 & 655  \\
\hline
         & $e\nu$    & 390 & 391 & 393  & 416   & 423 & 452 \\
  750    & $\mu\nu$  & 227 & 228 & 228  & 248   & 320 & 350 \\
         & both      & 186 & 184 & 188  & 208   & 232 & 259 \\
\hline
         & $e\nu$    & 199 & 200 & 200  & 207   & 217 & 225 \\
 1000    & $\mu\nu$  & 216 & 216 & 216  & 226   & 320 & 326 \\
         & both      & 108 & 109 & 109  & 115   & 133 & 141 \\
\hline
         & $e\nu$    & 149 & 150 & 150  & 153   & 163 & 167 \\
 1250    & $\mu\nu$  & 213 & 214 & 213  & 220   & 323 & 333 \\
         & both      &  88 &  88 &  88  &  91   & 108 & 112 \\
\hline
         & $e\nu$    & 155 & 156 & 156  & 159   & 169 & 173 \\
 1500    & $\mu\nu$  & 215 & 215 & 215  & 221   & 327 & 336 \\
         & both      &  90 &  90 &  90  &  93   & 111 & 115 \\
\hline
         & $e\nu$    & 164 & 163 & 164  & 168   & 171 & 175 \\
 1750    & $\mu\nu$  & 229 & 229 & 229  & 235   & 324 & 332 \\
         & both      &  95 &  96 &  96  &  98   & 112 & 115 \\
\hline
\hline
\end{tabular}
\end{center}
\end{table}



\begin{table}[!htbp]
\caption{Lower limits on \wp\ and \wstar\ masses.
         The first column is the decay channel ($e\nu$, $\mu\nu$ or both combined) and
         the following give the expected (Exp.) and observed (Obs.) mass limits.
\label{tab:limits_mass}
}
\begin{center}
\begin{tabular}{c|rr|rr}
\hline
\hline
           & \multicolumn{4}{c}{Mass limit [\gev]} \\
           & \multicolumn{2}{c|}{\wp} & \multicolumn{2}{c}{\wstar} \\
 decay     &  Exp. & Obs. & Exp. & Obs. \\
\hline
 $e\nu$    & 1370 & 1370 & 1260 & 1260 \\
 $\mu\nu$  & 1210 & 1290 & 1020 & 1120 \\
 both      & 1450 & 1490 & 1320 & 1350 \\
\hline
\hline
\end{tabular}
\end{center}
\end{table}

Limits on \wpl\ have been reported in many other
experiments~\cite{pdg:2010, d0:Wprime, cdf:Wprime, cdf:Wprime2010, cms:wprime2010, cms:wpmu2011mar}.
Prior to this letter and the recent \wpmu\ results from CMS~\cite{cms:wpmu2011mar},
the best limits in the high-mass region were reported
by CDF~\cite{cdf:Wprime2010} and CMS~\cite{cms:wprime2010}, both for \wpe.
The CDF measurement was made  with \ppbar\ collisions at $\sqrt{s}$~=~1.96~\tev\
using an integrated luminosity of~5.3~\ifb.
Both CMS results were obtained at the same collision energy ($\sqrt{s}$~=~7~\tev)
and during the same run period as those reported here.
The CMS limits were set using a Bayesian approach.
Ref.~\cite{cms:wpmu2011mar} also reports a combination of the CMS results in the two decay channels
with an SSM \wp\ mass limit of 1580~\gev.
Figure~\ref{fig:sigrat_comparison} compares the result presented here with the \wpe\ result from CDF
and the combination from CMS.
The comparison is made using the ratio of the limit to the calculated value of \xbr,
a quantity that is proportional to the square of the coupling strength.
The NNLO cross sections in Table~\ref{tab:xsec} are used for both the ATLAS and CMS points.

\begin{figure}[!htbp]
  \centering
  \includegraphics[width=0.49\textwidth]{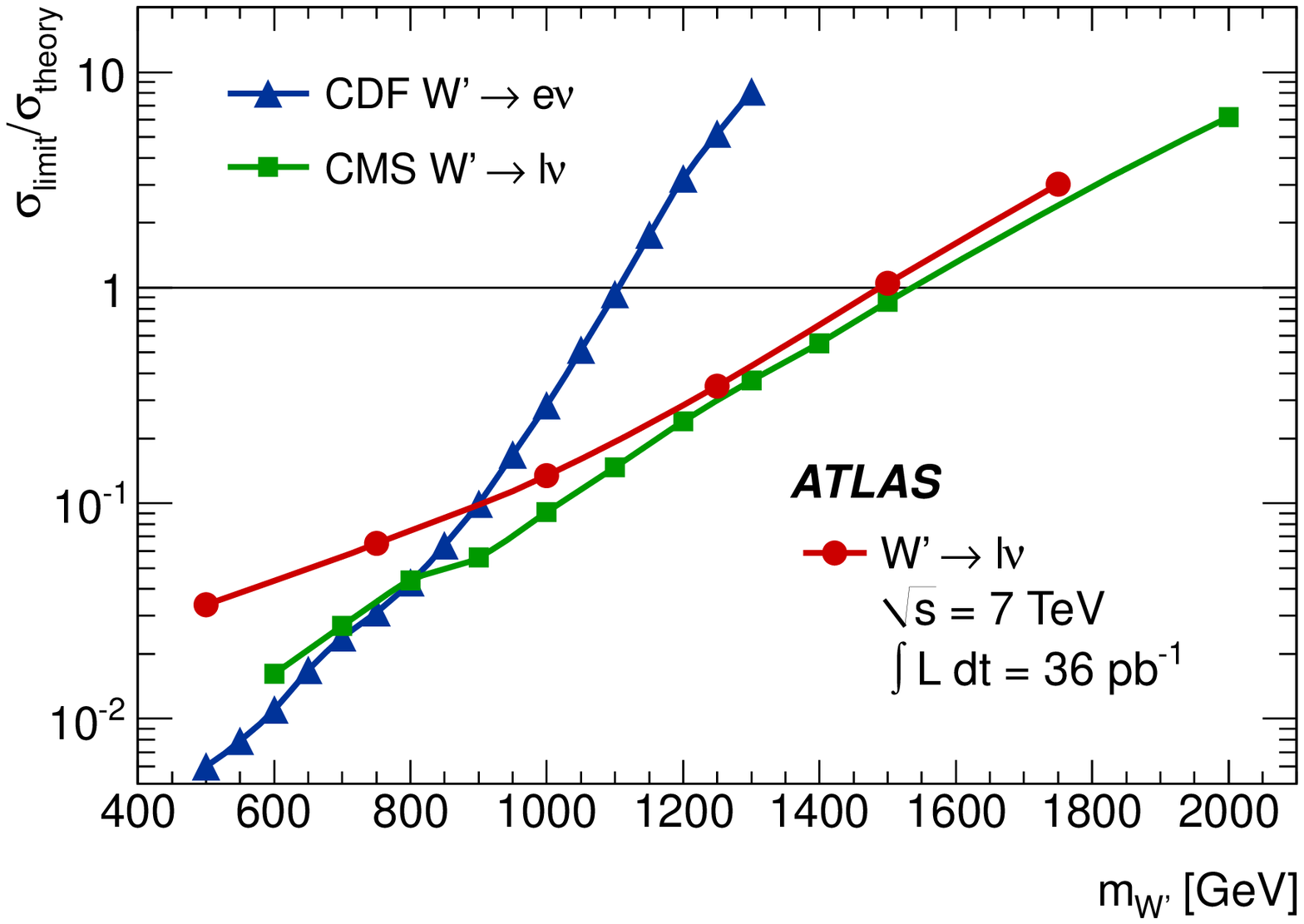}
  \caption{
  Normalized cross section limits ($\sigma_{\rm limit}/\sigma_{\rm theory}$) for \wp\
  as a function of mass for this measurement and those from CDF and CMS. The cross section
  calculations assume the \wp\ has the same couplings as the standard model \w~boson.
  The region above each curve is excluded at 95\% CL.
  \label{fig:sigrat_comparison}
  }
\end{figure}

In conclusion, the ATLAS detector has been used to search for new high-mass states decaying to a
lepton plus \mettext\ in \pp\ collisions at $\sqrt{s}=7\tev$  using 36~\ipb\ of integrated luminosity.
No excess beyond SM expectations is observed.
Limits on \xbr\ are shown in Figs.~\ref{fig:limits_xbr} and~\ref{fig:sigrat_comparison}.
A \wp\ with SSM couplings is excluded for masses below 1490~\gev\ at 95\% CL.
The exclusion for \wstar\ with couplings set in accordance with reference~\cite{wzstar_refmod2} is 1350~\gev.
The limits for \wstar\ are the most stringent to date.
   
\global\emergencystretch = .2\hsize



\section*{Acknowledgements}

We wish to thank CERN for the efficient commissioning and operation of the LHC during this initial high-energy data-taking period as well as the support staff from our institutions without whom ATLAS could not be operated efficiently.

We acknowledge the support of ANPCyT, Argentina; YerPhI, Armenia; ARC, Australia; BMWF, Austria; ANAS, Azerbaijan; SSTC, Belarus; CNPq and FAPESP, Brazil; NSERC, NRC and CFI, Canada; CERN; CONICYT, Chile; CAS, MOST and NSFC, China; COLCIENCIAS, Colombia; MSMT CR, MPO CR and VSC CR, Czech Republic; DNRF, DNSRC and Lundbeck Foundation, Denmark; ARTEMIS, European Union; IN2P3-CNRS, CEA-DSM/IRFU, France; GNAS, Georgia; BMBF, DFG, HGF, MPG and AvH Foundation, Germany; GSRT, Greece; ISF, MINERVA, GIF, DIP and Benoziyo Center, Israel; INFN, Italy; MEXT and JSPS, Japan; CNRST, Morocco; FOM and NWO, Netherlands; RCN, Norway;  MNiSW, Poland; GRICES and FCT, Portugal;  MERYS (MECTS), Romania;  MES of Russia and ROSATOM, Russian Federation; JINR; MSTD, Serbia; MSSR, Slovakia; ARRS and MVZT, Slovenia; DST/NRF, South Africa; MICINN, Spain; SRC and Wallenberg Foundation, Sweden; SER,  SNSF and Cantons of Bern and Geneva, Switzerland;  NSC, Taiwan; TAEK, Turkey; STFC, the Royal Society and Leverhulme Trust, United Kingdom; DOE and NSF, United States of America.  

The crucial computing support from all WLCG partners is acknowledged gratefully, in particular from CERN and the ATLAS Tier-1 facilities at TRIUMF (Canada), NDGF (Denmark, Norway, Sweden), CC-IN2P3 (France), KIT/GridKA (Germany), INFN-CNAF (Italy), NL-T1 (Netherlands), PIC (Spain), ASGC (Taiwan), RAL (UK) and BNL (USA) and in the Tier-2 facilities worldwide.


\bibliographystyle{model1-num-names}
\bibliography{plb}{}

\onecolumn
\begin{flushleft}
{\Large The ATLAS Collaboration}

\bigskip

G.~Aad$^{\rm 48}$,
B.~Abbott$^{\rm 112}$,
J.~Abdallah$^{\rm 11}$,
A.A.~Abdelalim$^{\rm 49}$,
A.~Abdesselam$^{\rm 119}$,
O.~Abdinov$^{\rm 10}$,
B.~Abi$^{\rm 113}$,
M.~Abolins$^{\rm 89}$,
H.~Abramowicz$^{\rm 154}$,
H.~Abreu$^{\rm 116}$
% ...
\bigskip

% ...
$^{10}$ Institute of Physics, Azerbaijan Academy of Sciences, Baku, Azerbaijan\\
$^{11}$ Institut de F\'isica d'Altes Energies and Universitat Aut\`onoma  de Barcelona and ICREA, Barcelona, Spain\\
% ...
$^{48}$ Fakult\"{a}t f\"{u}r Mathematik und Physik, Albert-Ludwigs-Universit\"{a}t, Freiburg i.Br., Germany\\
$^{49}$ Section de Physique, Universit\'e de Gen\`eve, Geneva, Switzerland\\
% ...
$^{89}$ Department of Physics and Astronomy, Michigan State University, East Lansing MI, United States of America\\
% ...
$^{112}$ Homer L. Dodge Department of Physics and Astronomy, University of Oklahoma, Norman OK, United States of America\\
$^{113}$ Department of Physics, Oklahoma State University, Stillwater OK, United States of America\\
% ...
$^{116}$ LAL, Univ. Paris-Sud and CNRS/IN2P3, Orsay, France\\
% ...
$^{119}$ Department of Physics, Oxford University, Oxford, United Kingdom\\
% ...
$^{154}$ Raymond and Beverly Sackler School of Physics and Astronomy, Tel Aviv University, Tel Aviv, Israel\\
% ...
\end{flushleft}

\end{document}